\documentclass{article}

\usepackage{arxiv}

\usepackage[utf8]{inputenc} 
\usepackage[T1]{fontenc}    
\usepackage{hyperref}       
\usepackage{url}            
\usepackage{booktabs}       
\usepackage{amsfonts}       
\usepackage{nicefrac}       
\usepackage{microtype}      
\usepackage{lipsum}
\usepackage{graphicx}

\usepackage{cite}
\usepackage{tabularx}
\usepackage{amsmath,amssymb,amsfonts}
\usepackage{algorithmic}
\usepackage{adjustbox}
\usepackage{textcomp}
\usepackage{ragged2e}
\usepackage{marvosym}
\usepackage{subcaption}
\usepackage{float}

\graphicspath{ {./images/} }

\title{A Decentralized Framework for Ethical Authorship Validation in Academic Publishing: Leveraging Self-Sovereign Identity and Blockchain Technology}

\author{
Kamal Al-Sabahi \\
College of Banking and Financial Studies\\
Muscat, Sultanate of Oman \\
\texttt{firstname@cbfs.edu.cn} \\
\And
Yousuf Khamis Al Mabsali \\
College of Banking and Financial Studies\\
Muscat, Sultanate of Oman \\
}

\begin{document}
\maketitle

\begin{abstract}
Academic publishing, integral to knowledge dissemination and scientific advancement, increasingly faces threats from unethical practices such as unconsented authorship, gift authorship, author ambiguity, and undisclosed conflicts of interest. While existing infrastructures like ORCID effectively disambiguate researcher identities, they fall short in enforcing explicit authorship consent, accurately verifying contributor roles, and robustly detecting conflicts of interest during peer review. To address these shortcomings, this paper introduces a decentralized framework leveraging Self-Sovereign Identity (SSI) and blockchain technology. The proposed model uses Decentralized Identifiers (DIDs) and Verifiable Credentials (VCs) to securely verify author identities and contributions, reducing ambiguity and ensuring accurate attribution. A blockchain-based trust registry records authorship consent and peer-review activity immutably. Privacy-preserving cryptographic techniques, especially Zero-Knowledge Proofs (ZKPs), support conflict-of-interest detection without revealing sensitive data. Verified authorship metadata and consent records are embedded in publications, increasing transparency. A stakeholder survey of researchers, editors, and reviewers suggests the framework improves ethical compliance and confidence in scholarly communication. This work represents a step toward a more transparent, accountable, and trustworthy academic publishing ecosystem.
\end{abstract}


\section{Introduction}\label{sec:introduction}

Academic publishing serves as an essential cornerstone in the dissemination and advancement of scientific knowledge, fostering cross-disciplinary collaboration and innovation. Despite its critical role, the integrity of this system is increasingly compromised by unethical practices, severely threatening credibility and reliability. The magnitude of this issue was exemplified in 2023, with more than 10,000 research papers retracted—a historical peak—signaling a troubling rise in questionable publications~\cite{van2023more}. Integrity experts caution that these retractions likely represent merely the tip of the iceberg, reflecting profound systemic flaws inherent within scholarly communication. Specifically, unethical practices such as unconsented authorship, exclusion of legitimate contributors, and misuse of gift authorship to exploit prominent individuals' reputations have significantly compromised the authenticity of scholarly outputs~\cite{johann2022perceptions, shubha2021publication}. Furthermore, unresolved conflicts of interest and cases of impersonation within the peer-review process have further eroded trust among academic communities~\cite{nitin2020governing}.

To address these growing concerns, prior studies have explored blockchain technologies as potential solutions within academic publishing~\cite{lukas2021blockchain, rafael2020ssibac}. However, these efforts often overlook critical ethical considerations such as explicit authorship consent and rigorous conflict management. Existing identity management solutions, including widely-adopted platforms like ORCID, successfully disambiguate author identities through unique identifiers. Nevertheless, these systems fall short in adequately verifying authorship roles, securing explicit consent, and detecting conflicts of interest during peer review~\cite{cristian2024design, alex2021selfsovereign}. Their limitations have inadvertently enabled predatory journals to fabricate authorship and editorial credentials, further exacerbating the need for a robust, secure, and transparent framework for ethical authorship validation~\cite{lukas2021blockchain}.

In response to these critical challenges, this paper proposes a novel decentralized framework leveraging Self-Sovereign Identity (SSI) and blockchain technology. Specifically, Decentralized Identifiers (DIDs) and Verifiable Credentials (VCs) are utilized to establish cryptographically secure and verifiable author identities and contribution records~\cite{stefano2023survey}. The proposed system integrates blockchain as an immutable trust registry, guaranteeing tamper-proof documentation of authorship consent and metadata. Additionally, advanced privacy-preserving cryptographic techniques, such as Zero-Knowledge Proofs (ZKPs) and Selective Disclosure, are employed to securely manage conflict-of-interest detection without compromising sensitive data~\cite{nitin2020governing, efat2022dtssim}.

The key contributions of this work can be summarized as follows:

\begin{enumerate}
    \item \textbf{Development of a Decentralized Framework:} A secure and decentralized system is proposed, utilizing Self-Sovereign Identity (SSI) and Verifiable Credentials (VCs) to robustly validate authorship, ensure cryptographically verifiable identities, and maintain tamper-proof records.
    
    \item \textbf{Privacy-Preserving Conflict Detection:} The framework introduces Zero-Knowledge Proofs (ZKPs) and Selective Disclosure methods to effectively detect conflicts of interest while preserving user privacy and confidentiality.
    
    \item \textbf{Immutable Authorship Consent Records:} A blockchain-based process is established to ensure immutable, auditable records of authorship consent and verified metadata, significantly enhancing transparency and accountability in academic publishing.
    
    \item \textbf{Comprehensive Stakeholder Evaluation:} A detailed survey was conducted involving researchers, editors, and publishers to evaluate the proposed framework's feasibility, effectiveness, and acceptance, providing valuable insights into existing ethical challenges and the system's perceived utility.
\end{enumerate}

The remainder of this paper is organized as follows: Section~\ref{sec:related_work} reviews related work on blockchain applications, authorship validation, and conflict management within academic publishing. Section~\ref{sec:background} introduces core concepts underpinning the proposed framework, including SSI, blockchain technology, and cryptographic techniques. Section~\ref{sec:framework} details the proposed model’s architecture, components, and operational workflow. Section~\ref{sec:evaluation} provides a comprehensive survey-based evaluation, highlighting critical findings and stakeholder insights. Finally, Section~\ref{sec:conclusion} concludes the paper by summarizing the primary contributions, addressing limitations, and proposing future research directions.

\section{Related Work}\label{sec:related_work}

The rapid evolution of academic publishing necessitates the integration of emerging technologies to address ethical, procedural, and systemic challenges inherent to scholarly communication~\cite{mohd2021blockchain}. Recently, blockchain technology and Self-Sovereign Identity (SSI) have garnered significant attention due to their potential in enhancing transparency, accountability, and efficiency across academic workflows. This section reviews existing literature relevant to blockchain applications in publishing, authorship validation systems, and peer review mechanisms, highlighting their contributions as well as existing limitations.

\subsection{Blockchain Applications in Academic Publishing}

Blockchain technology offers unique advantages such as immutability, decentralization, and transparency, positioning it effectively to tackle persistent challenges in academic publishing. Petr et al.~\cite{petr2018permissioned} explored permissioned blockchains, specifically Hyperledger Fabric, to enhance transparency in research processes, including productivity tracking, reputation management, and peer review. Similarly, Tarkhanov et al.~\cite{tarkhanov2019application} utilized the Ethereum blockchain to guarantee data immutability in online scientific journals, demonstrating blockchain's capability to verify publication integrity.

Mohan~\cite{mohan2019blockchain} examined blockchain as a remedy to academic misconduct, proposing a permissioned blockchain-based framework aimed at mitigating the ``Academic Dilemma," a situation where competitive pressures lead researchers toward unethical practices. Expanding upon this concept, Mackey et al.~\cite{mackey2019framework} proposed a consortium blockchain model employing Democratic Autonomous Organizations (DAOs) to oversee submission, peer review, and editorial decision-making, thereby democratizing traditional academic publishing processes.

Moreover, Tenorio-Fornés et al.~\cite{tenorio2019decentralized} developed a decentralized peer-review and publication system combining blockchain and the InterPlanetary File System (IPFS). Their system incorporated distributed reviewer reputation mechanisms alongside open-access infrastructures, enhancing transparency. Nevertheless, these studies often neglect critical challenges, such as privacy management, scalability, and resistance to adoption within academic communities.

\subsection{Authorship Validation and Identity Management}

Authorship validation remains a significant concern in scholarly communication. Traditional identity management solutions, exemplified by platforms such as ORCID~\cite{laurel2012orcid}, primarily focus on disambiguating author identities through unique identifiers and integrating metadata with systems like CrossRef. However, these platforms fall short in enforcing explicit authorship consent or preventing unethical authorship practices, such as gift authorship or ghost authorship.

Emerging blockchain-based systems offer innovative solutions to address these limitations by providing tamper-proof authorship records and streamlined workflows. For instance, Tenorio-Fornés et al.~\cite{tenorio2019decentralized} highlighted decentralized identity frameworks to enhance transparency in publication processes. Nevertheless, these approaches frequently prioritize transparency over privacy, failing to effectively safeguard sensitive identity-related information. Our proposed model addresses this critical gap by leveraging SSI principles, employing Decentralized Identifiers (DIDs), and Verifiable Credentials (VCs) to ensure secure, consent-driven, and privacy-preserving authorship validation.

\subsection{Peer Review Transparency and Associated Challenges}

Peer review is a cornerstone of academic publishing but is frequently criticized for lacking transparency, being vulnerable to biases, and suffering inefficiencies. Tennant et al.~\cite{tennant2018peerreview} identified critical challenges including editorial responsibility, reviewer subjectivity, and inconsistency in review standards, calling for standardized procedures and shared data infrastructures. Similarly, Jonathan et al.~\cite{jonathan2020limitations} advocated enhanced transparency and accountability to bolster peer review legitimacy.

Blockchain-enabled approaches offer promising remedies to these issues. Mackey et al.~\cite{mackey2019framework} proposed leveraging DAOs to manage peer-review activities, generating immutable records of reviewer interactions. Additionally, Tenorio-Fornés et al.~\cite{tenorio2019decentralized} introduced reputation mechanisms combined with privacy-preserving techniques to incentivize transparent, high-quality peer reviews. Nevertheless, widespread adoption of blockchain-based peer-review systems remains hindered by technical complexity and resistance from stakeholders accustomed to traditional workflows~\cite{jesse2016where}.

\subsection{Broader Blockchain Applications and Lessons Learned}

The versatility of blockchain extends beyond academic publishing into various domains, offering valuable insights applicable to scholarly communication. Yin et al.~\cite{yin2023blockchain} discussed blockchain applications in corporate governance, emphasizing benefits like transparency, real-time accessibility, and cost-effective record management. In healthcare, blockchain frameworks such as those proposed by Gaby et al.~\cite{gaby2018ancile} aim to balance data privacy and accessibility in electronic health records management. Despite demonstrating promising capabilities, blockchain applications in these sectors also highlight consistent challenges related to scalability, interoperability, and user adoption—challenges equally pertinent to academic publishing workflows.

\subsection{Gaps and Research Opportunities}

Despite significant advancements in applying blockchain to academic publishing, existing solutions continue to exhibit notable limitations, summarized as follows:

\begin{itemize}
    \item \textbf{Ethical Oversight:} Existing platforms such as ORCID do not enforce explicit authorship consent nor adequately address unethical authorship practices~\cite{laurel2012orcid}.
    
    \item \textbf{Privacy Protection:} Current blockchain-based solutions often insufficiently protect sensitive identity information during authorship validation and peer review processes, raising privacy concerns among stakeholders~\cite{tenorio2019decentralized}.
    
    \item \textbf{Scalability and Interoperability:} Integration of decentralized blockchain frameworks with established platforms like CrossRef and Scopus requires significant technical collaboration and adherence to standardized protocols~\cite{mackey2019framework}.
    
    \item \textbf{Resistance to Adoption:} Transitioning stakeholders from traditional centralized systems toward decentralized blockchain infrastructures demands considerable educational efforts and cultural adaptation~\cite{jesse2016where}.
\end{itemize}

The model proposed in this paper directly addresses these gaps, combining blockchain and SSI technologies to achieve tamper-proof authorship consent records, robust conflict-of-interest detection via privacy-preserving cryptographic techniques, and interoperable metadata embedding. By explicitly tackling ethical and technical challenges, this work provides a robust pathway toward enhanced transparency, accountability, and trust within academic publishing.

\section{Preliminaries}\label{sec:background}

This section provides foundational concepts integral to the proposed decentralized authorship validation framework, including Self-Sovereign Identity (SSI), blockchain technology, and advanced cryptographic techniques. These components collectively address core challenges of authorship validation, consent enforcement, and privacy preservation within academic publishing.

\subsection{Self-Sovereign Identity (SSI)}

Digital identity broadly encompasses attributes and identifiers associated with an individual or entity within a specific digital context, facilitating authentication and authorization processes online~\cite{lux2020distributed}. In contrast to conventional identity management systems, which typically rely on centralized service providers, Self-Sovereign Identity (SSI) represents a paradigm shift by enabling individuals to maintain control and ownership over their digital identities~\cite{Pava2023bibliometric}. SSI empowers users with sovereign, persistent, and portable identities, providing secure access to digital services without reliance on intermediaries, thereby significantly enhancing user privacy and security~\cite{alex2021selfsovereign, Schardong2022self}.

\begin{figure*}[htbp]
    \centering
    \includegraphics[width=0.8\textwidth]{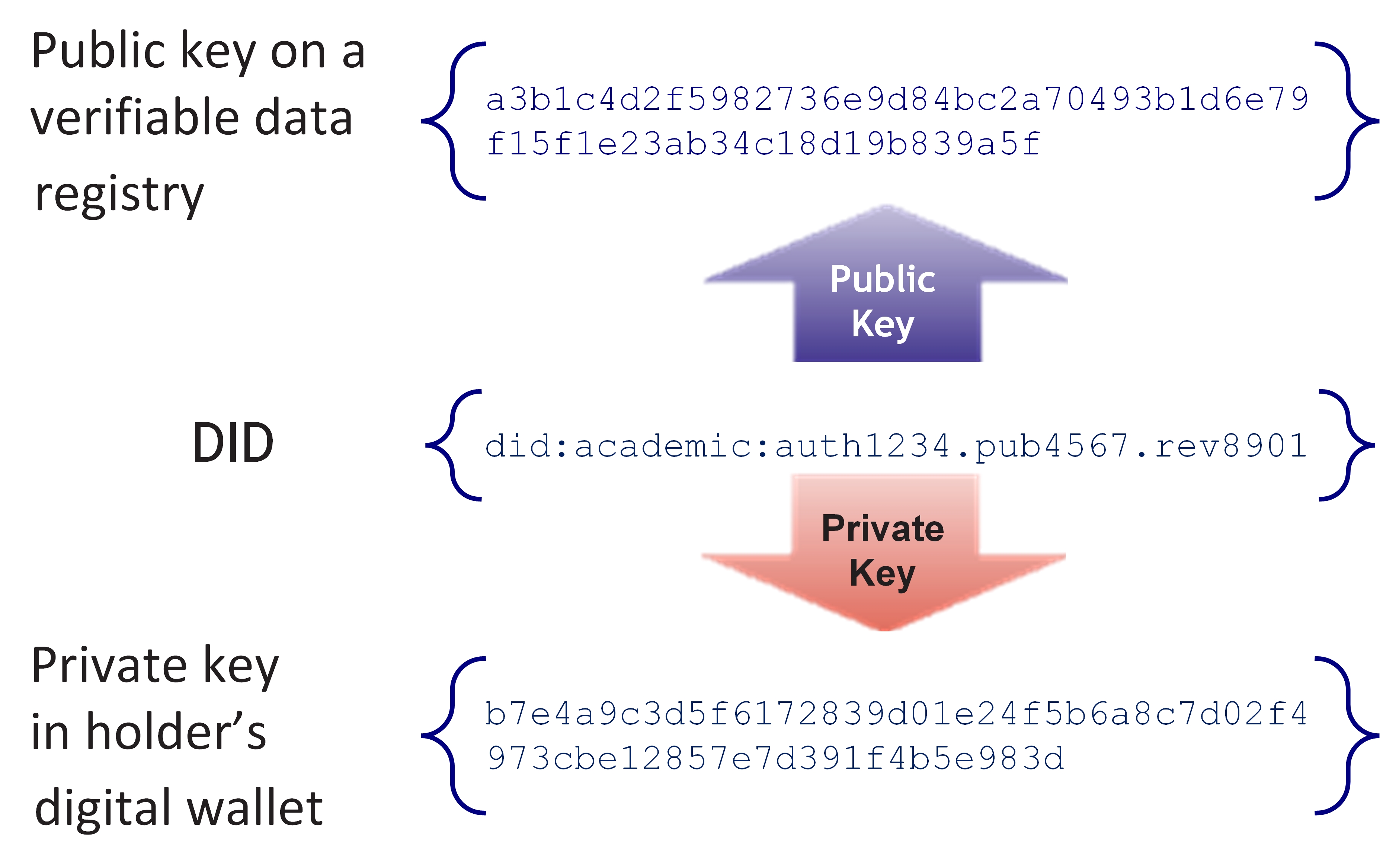}
    \caption{Example of a decentralized identifier (DID) structure illustrating the association between public and private cryptographic keys. The DID links to a publicly accessible key stored within a decentralized registry, while the private key remains securely managed by the identity holder to ensure secure cryptographic interactions.}
    \label{fig:did_example}
\end{figure*}

Central to the SSI architecture are Decentralized Identifiers (DIDs), cryptographically secure, globally unique identifiers that enable autonomous identity management independent of centralized authorities~\cite{stefano2023survey}. Figure~\ref{fig:did_example} illustrates a representative DID structure, showing how the public key associated with the DID is stored in a decentralized registry (e.g., blockchain), while the corresponding private key remains securely held by the individual user within a digital wallet, facilitating cryptographic verification during interactions.

Complementing DIDs are Verifiable Credentials (VCs), digitally signed attestations issued by trusted entities to validate specific user attributes such as institutional affiliation, credentials, and qualifications~\cite{lux2020distributed}. VCs allow selective disclosure of identity attributes, enabling users to provide only the minimal information required for verification, thereby preserving data privacy.

\begin{figure*}[htbp]
\centering
\includegraphics[width=\textwidth]{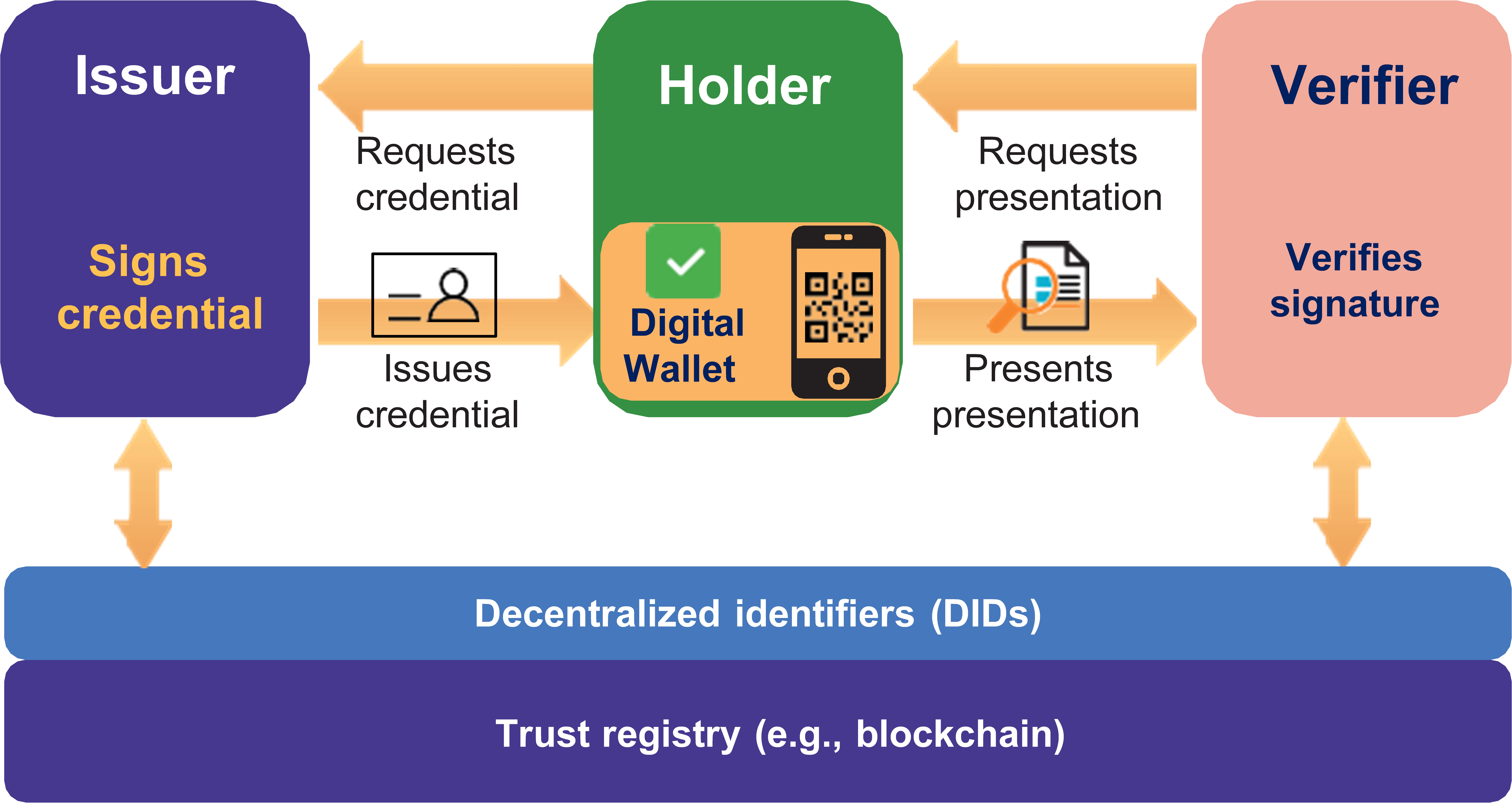}
\caption{Core interactions within an SSI framework: The Issuer creates and signs Verifiable Credentials (VCs); the Holder securely manages and selectively shares these credentials; and the Verifier cryptographically validates credentials using public keys stored on a decentralized blockchain-based trust registry.}
\label{fig:ssi_roles}
\end{figure*}

The core interactions within the SSI framework are depicted in Figure~\ref{fig:ssi_roles}. The primary stakeholders include: (i) the \textit{Issuer}, who generates and cryptographically signs credentials; (ii) the \textit{Holder}, who stores credentials securely in a digital wallet; and (iii) the \textit{Verifier}, who requests and cryptographically validates credentials presented by the holder. The blockchain acts as a decentralized, tamper-proof trust registry, facilitating secure storage and verification of DIDs and credential hashes, thereby reinforcing privacy and security.

In the context of academic publishing, SSI ensures secure, verifiable author identities and affiliations, addresses ambiguity and consent challenges, and safeguards sensitive personal information during authorship and peer review processes~\cite{lukas2021blockchain, rafael2020ssibac}.

\subsection{Blockchain Technology}

Blockchain technology serves as a foundational infrastructure for SSI by providing a decentralized, transparent, and immutable ledger for securely recording cryptographic hashes of identity-related information, including DIDs and VCs~\cite{soltani2021survey}. By storing only cryptographic hashes rather than sensitive data directly, blockchain ensures privacy protection while enabling robust, transparent verification processes. The key characteristics of blockchain relevant to authorship validation and SSI frameworks include:

\begin{itemize}
    \item \textbf{Immutability:} Data entries recorded on the blockchain ledger are irreversible and tamper-proof, preserving the integrity of authorship records and consent documentation~\cite{nitin2020governing, cristian2024design}.
    
    \item \textbf{Transparency:} Blockchain enables authorized stakeholders to independently verify recorded information without exposing confidential details, thereby enhancing accountability and fostering trust among authors, reviewers, and publishers~\cite{efat2022dtssim}.
    
    \item \textbf{Decentralization:} The distributed architecture of blockchain eliminates reliance on centralized authorities, thereby reducing risks of manipulation, fraud, and single points of failure, ensuring robust security and reliability~\cite{yirui2022decentralized}.
\end{itemize}

Moreover, blockchain infrastructures facilitate advanced cryptographic methods like Zero-Knowledge Proofs (ZKPs), further reinforcing privacy-preserving verification capabilities~\cite{stefano2023survey, alex2021selfsovereign}.

\subsection{Cryptographic Techniques}

Advanced cryptographic techniques are integral to SSI systems, ensuring data security, privacy, and compliance with ethical standards. Two notable cryptographic mechanisms leveraged within this framework include:

\begin{enumerate}
    \item \textbf{Selective Disclosure:} Selective disclosure enables individuals to selectively reveal certain credential attributes necessary for specific verifications, withholding non-essential personal data. For instance, an author can verify institutional affiliation without disclosing additional sensitive personal details, thus preserving privacy~\cite{lukas2021blockchain, nitin2020governing}.
    
    \item \textbf{Zero-Knowledge Proofs (ZKPs):} ZKPs allow individuals to cryptographically prove specific claims (e.g., the absence of conflicts of interest) without revealing the underlying details or data supporting those claims. This approach effectively safeguards privacy while ensuring compliance and ethical standards in peer review and authorship verification~\cite{efat2022dtssim, rafael2020ssibac}.
\end{enumerate}

\subsection{Persistent Authorship Challenges in Academic Publishing}

Academic publishing faces several persistent ethical challenges concerning authorship validation, including:

\begin{itemize}
    \item \textbf{Unconsented Authorship:} Inclusion of individuals as authors without their explicit approval.
    
    \item \textbf{Gift Authorship:} Adding authors with prominent reputations who have made no substantial contributions, to enhance manuscript credibility.
    
    \item \textbf{Ghost Authorship:} Exclusion of rightful contributors from author lists, diminishing credit allocation.
    
    \item \textbf{Conflict of Interest in Peer Review:} Reviewers with undisclosed conflicts of interest, compromising peer-review objectivity and integrity.
\end{itemize}

Although existing platforms, such as ORCID, provide mechanisms for researcher identification, they lack capabilities for enforcing explicit authorship consent, reliably detecting conflicts of interest, and creating immutable records of verified contributions. These limitations exacerbate accountability gaps and ethical risks within academic publishing processes~\cite{yirui2022decentralized, efat2022dtssim}.
\section{Proposed Model}\label{sec:framework}

\begin{figure*}[htbp]
    \centering
    \includegraphics[width=\textwidth]{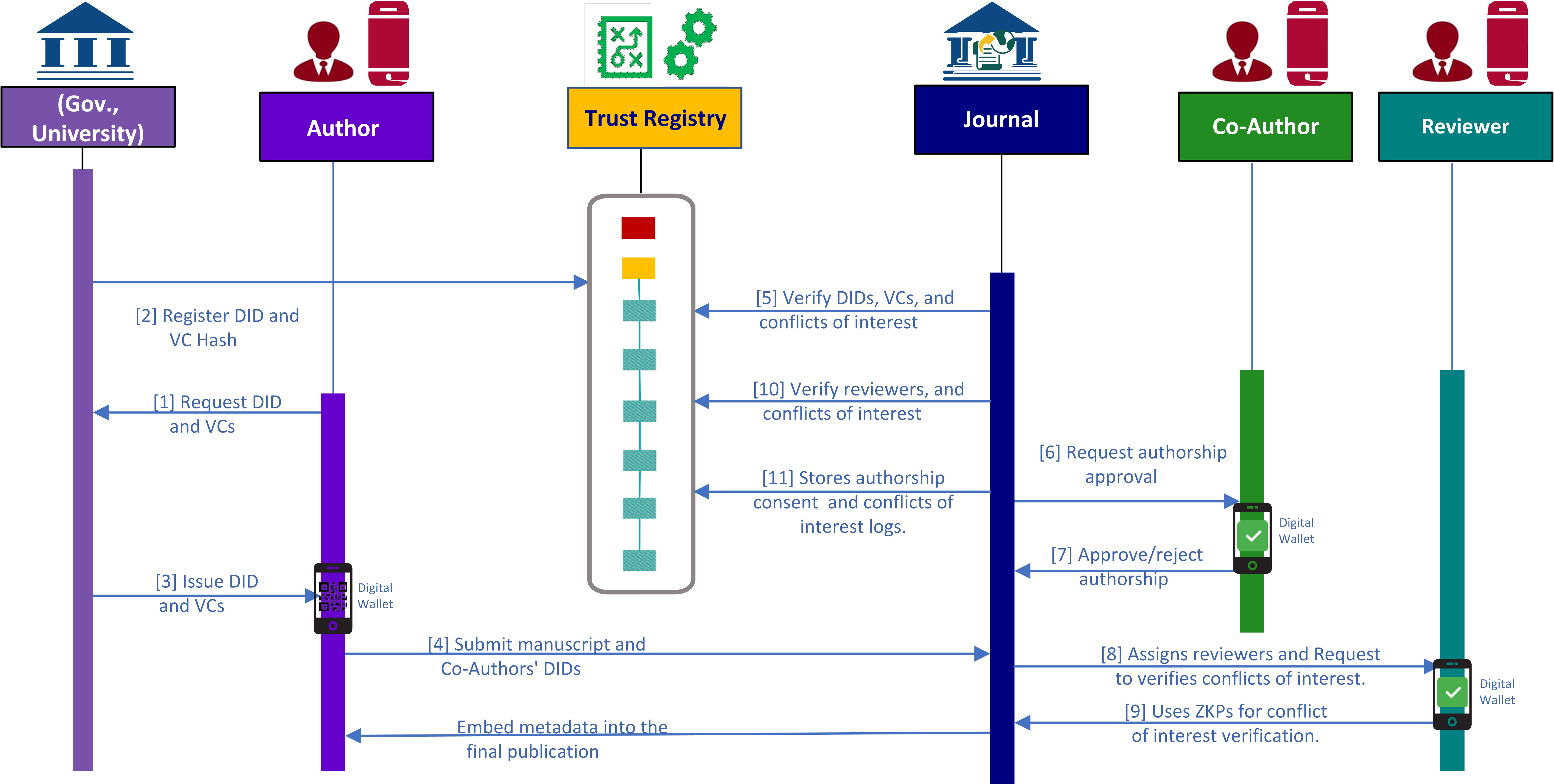}
    \caption{Architecture and workflow of the proposed decentralized authorship validation framework, depicting interactions among authors, co-authors, reviewers, journals, and trusted institutions. Core components include Decentralized Identifiers (DIDs), Verifiable Credentials (VCs), blockchain as a trust registry, and cryptographic techniques such as Zero-Knowledge Proofs (ZKPs) for privacy-preserving conflict-of-interest detection.}
    \label{fig:architecture}
\end{figure*}

In this section, we present a novel decentralized framework leveraging Self-Sovereign Identity (SSI) and blockchain technologies to comprehensively address prevalent challenges in academic publishing, such as unconsented authorship, impersonation, and undisclosed conflicts of interest. The proposed model ensures transparency, accountability, and ethical compliance, while rigorously safeguarding sensitive personal information. Figure~\ref{fig:architecture} illustrates the complete system architecture and workflow, highlighting interactions among key stakeholders: authors, co-authors, reviewers, journals, and trusted institutions.

\subsection{Architecture and Core Components}

The proposed framework integrates critical stakeholders—including authors, co-authors, reviewers, journals, and credential-issuing institutions—within a decentralized, blockchain-based infrastructure. This architecture ensures data immutability, privacy protection, and verifiable authorship claims. Specifically, the system consists of the following core components:

\begin{itemize}
    \item \textbf{Decentralized Identifiers (DIDs):} Cryptographically secure, unique identifiers assigned to authors, co-authors, and reviewers, providing robust authentication to prevent impersonation and ensuring trustworthy author attribution.
    
    \item \textbf{Verifiable Credentials (VCs):} Digitally signed attestations from authorized entities confirming users' qualifications, institutional affiliations, and specific research contributions. VCs enable selective disclosure, facilitating privacy-preserving verifications.
    
    \item \textbf{Blockchain Trust Registry:} A decentralized ledger maintaining cryptographic hashes of DIDs and VCs, ensuring data immutability and enabling transparent verification of authorship and reviewer credentials.
    
    \item \textbf{Journals as Coordinators:} Academic journals manage manuscript submission processes, verify author credentials via the blockchain, assign qualified reviewers, and oversee peer-review procedures, acting as intermediaries among stakeholders.
    
    \item \textbf{Privacy-Preserving Reviewer Verification:} Reviewers leverage cryptographic methods, such as Zero-Knowledge Proofs (ZKPs), to confirm the absence of conflicts of interest without disclosing private or sensitive details, thus preserving ethical standards and reviewer anonymity.
    
    \item \textbf{Transparent Authorship Metadata:} Authorship contributions, consent records, and verified credentials are timestamped and embedded directly into publication metadata, thereby enhancing traceability, accountability, and trust across the publication lifecycle.
\end{itemize}

\subsection{Operational Workflow}

The operational workflow of the proposed framework is structured into four clearly defined phases, seamlessly integrating the core components described above:

\begin{enumerate}
    \item \textbf{Identity Creation and Credential Issuance:} 
    Authors and reviewers initiate requests for Decentralized Identifiers (DIDs) and Verifiable Credentials (VCs) from trusted issuing institutions. Upon identity verification, these credentials are cryptographically signed, registered on the blockchain ledger, and securely stored in the individual's digital wallet for subsequent verification processes.

    \item \textbf{Manuscript Submission and Authorship Validation:} 
    During manuscript submission, primary authors provide their DIDs alongside those of co-authors within the submission metadata. Journals access the blockchain registry to validate these credentials and verify authorship roles. Verified co-authors are then prompted to explicitly consent to their authorship via digital signatures. Any absence or rejection of consent triggers an immediate alert for journal editorial resolution.

    \item \textbf{Peer Review and Conflict-of-Interest Detection:} 
    Journals assign peer reviewers based on verified credentials and areas of expertise recorded in their DIDs and VCs. Assigned reviewers utilize Zero-Knowledge Proofs (ZKPs) to cryptographically demonstrate the absence of conflicts of interest, ensuring confidentiality of sensitive relationships or affiliations. Only upon successful conflict verification do reviewers proceed to assess manuscript submissions.

    \item \textbf{Metadata Embedding and Publication:} 
    After successful peer review and editorial approval, journals embed verified authorship metadata—comprising authors' DIDs, validated contributions, roles, and consent records—into the published article. This metadata is immutably recorded on the blockchain, ensuring traceability and authenticity. Indexed databases and individual readers can independently verify publication integrity through blockchain-validated metadata.
\end{enumerate}

\subsection{Benefits of the Proposed Framework}

The proposed decentralized framework delivers multiple significant advantages, addressing critical ethical and procedural concerns in contemporary academic publishing:

\begin{itemize}
    \item \textbf{Enhanced Trust and Transparency:} By systematically addressing unethical authorship practices and conflicts of interest, the framework significantly enhances stakeholder trust, transparency, and confidence in scholarly outputs.
    
    \item \textbf{Privacy Preservation:} Robust cryptographic techniques, including Zero-Knowledge Proofs (ZKPs) and selective disclosure, effectively safeguard sensitive personal information while ensuring compliance with ethical standards.
    
    \item \textbf{Decentralization and Scalability:} The blockchain-based decentralized infrastructure eliminates single points of failure, mitigates risks of manipulation, and offers scalable solutions adaptable to diverse publishing environments and stakeholder needs.
    
    \item \textbf{Improved Accountability and Auditability:} Verified and timestamped authorship metadata permanently embedded into publications establishes clear accountability pathways, enabling retrospective audits and promoting ethical responsibility throughout academic publishing processes.
\end{itemize}

\section{Discussion and Evaluation}\label{sec:evaluation}

In this section, we evaluate the feasibility, effectiveness, and broader implications of the proposed decentralized Self-Sovereign Identity (SSI) framework for authorship validation. Specifically, we analyze insights obtained from a stakeholder survey, discuss implications for academic publishing practices, and identify limitations that must be addressed to ensure successful implementation.

\subsection{Survey Design}

To comprehensively evaluate the proposed framework's utility and relevance, we conducted a detailed stakeholder survey. The survey collected demographic and professional background data—including respondent roles and expertise—to contextualize responses accurately. Survey questions targeted stakeholders' experiences with unethical authorship practices, effectiveness of existing validation mechanisms (e.g., ORCID), and their perceptions of the proposed model’s potential impact. Questions were structured to capture quantitative data (e.g., Likert-scale and multiple-choice questions) alongside qualitative insights through open-ended questions.

Specifically, participants reported their experiences with unethical authorship practices such as unconsented authorship, gift authorship, ghost authorship, and exclusion of rightful contributors. Additionally, respondents evaluated the effectiveness of existing systems in enforcing authorship consent and preventing unethical practices. To measure perceived utility, participants were introduced to the framework’s core features, including blockchain-based tamper-proof records, privacy-preserving conflict detection via Zero-Knowledge Proofs (ZKPs), and transparent metadata embedding. Finally, the survey addressed potential barriers to adoption, focusing on privacy concerns, technical complexity, and integration challenges with existing platforms.

\subsection{Key Findings}

The survey outcomes provided valuable insights into current authorship validation practices and stakeholder perceptions, emphasizing the need for more robust systems. Key findings and their implications are summarized below.

\begin{figure}[htbp]
    \centering
    \begin{subfigure}[b]{0.45\textwidth}
        \includegraphics[width=\textwidth]{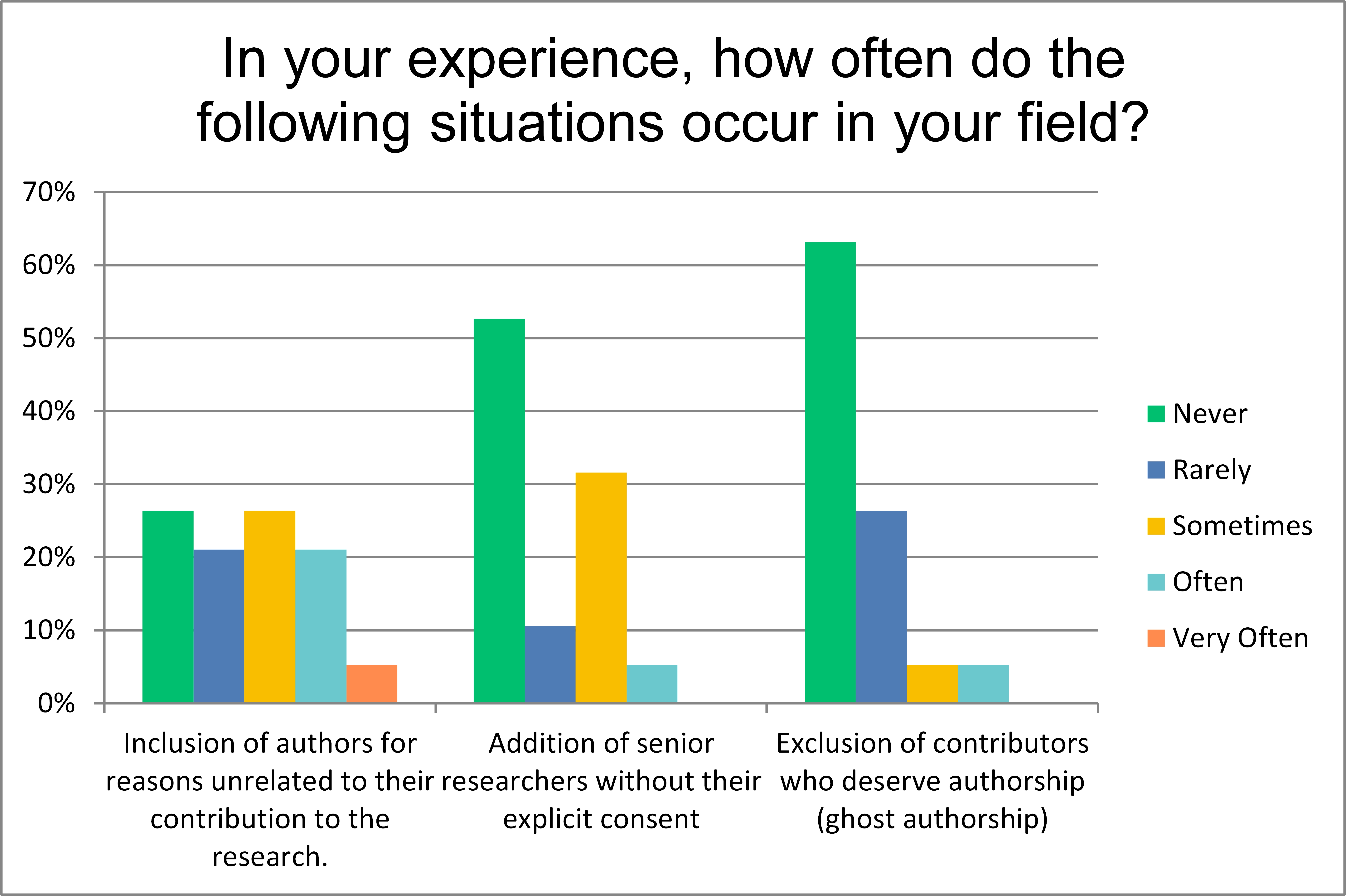}
        \caption{Frequency of unethical authorship practices.}
        \label{fig:q6}
    \end{subfigure}
    \hfill
    \begin{subfigure}[b]{0.45\textwidth}
        \includegraphics[width=\textwidth]{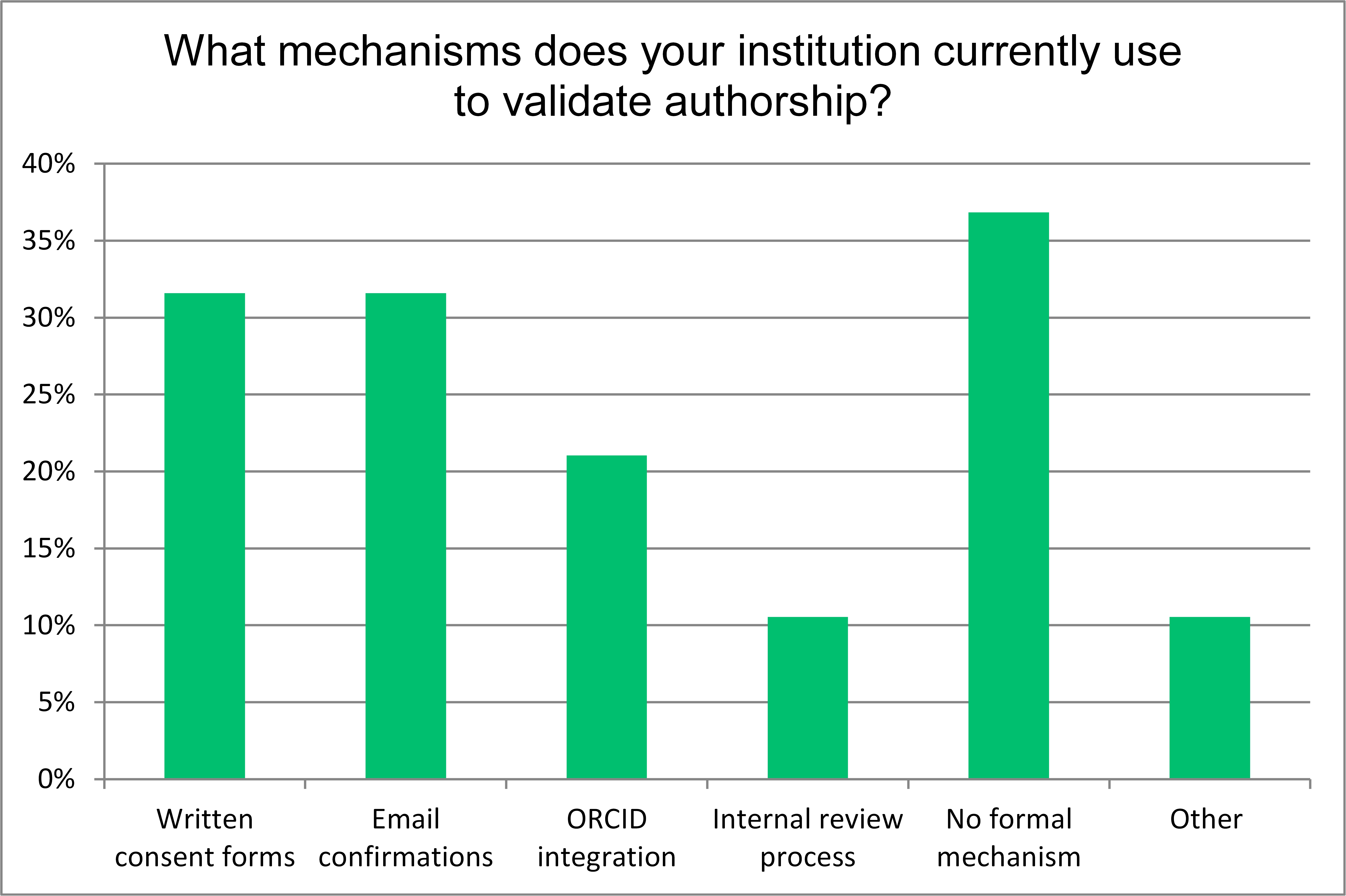}
        \caption{Institutional mechanisms for authorship validation.}
        \label{fig:q9}
    \end{subfigure}
    \caption{Survey results highlighting (a) the frequency of unethical authorship practices encountered, and (b) prevalent institutional mechanisms for authorship validation.}
    \label{fig:group1}
\end{figure}

\paragraph{Prevalence of Unethical Authorship Practices}
The results indicate widespread occurrences of unethical authorship practices: 47\% of respondents reported encountering unconsented authorship or gift authorship at least occasionally, while 36\% noted exclusion of deserving contributors (Figure~\ref{fig:q6}). This underscores a systemic integrity challenge in academic publishing, highlighting the urgency of mechanisms to effectively enforce accountability.

\paragraph{Current Authorship Validation Mechanisms}
Institutional approaches to authorship validation remain largely informal: 37\% of respondents indicated a complete lack of formal validation mechanisms, with reliance on informal methods such as written consent (32\%) and email confirmations (32\%) being common (Figure~\ref{fig:q9}). Notably, the adoption of structured, integrated systems like ORCID is limited (21\%), emphasizing the inadequacies of current methodologies.

\begin{figure}[htbp]
    \centering
    \begin{subfigure}[b]{0.45\textwidth}
        \includegraphics[width=\textwidth]{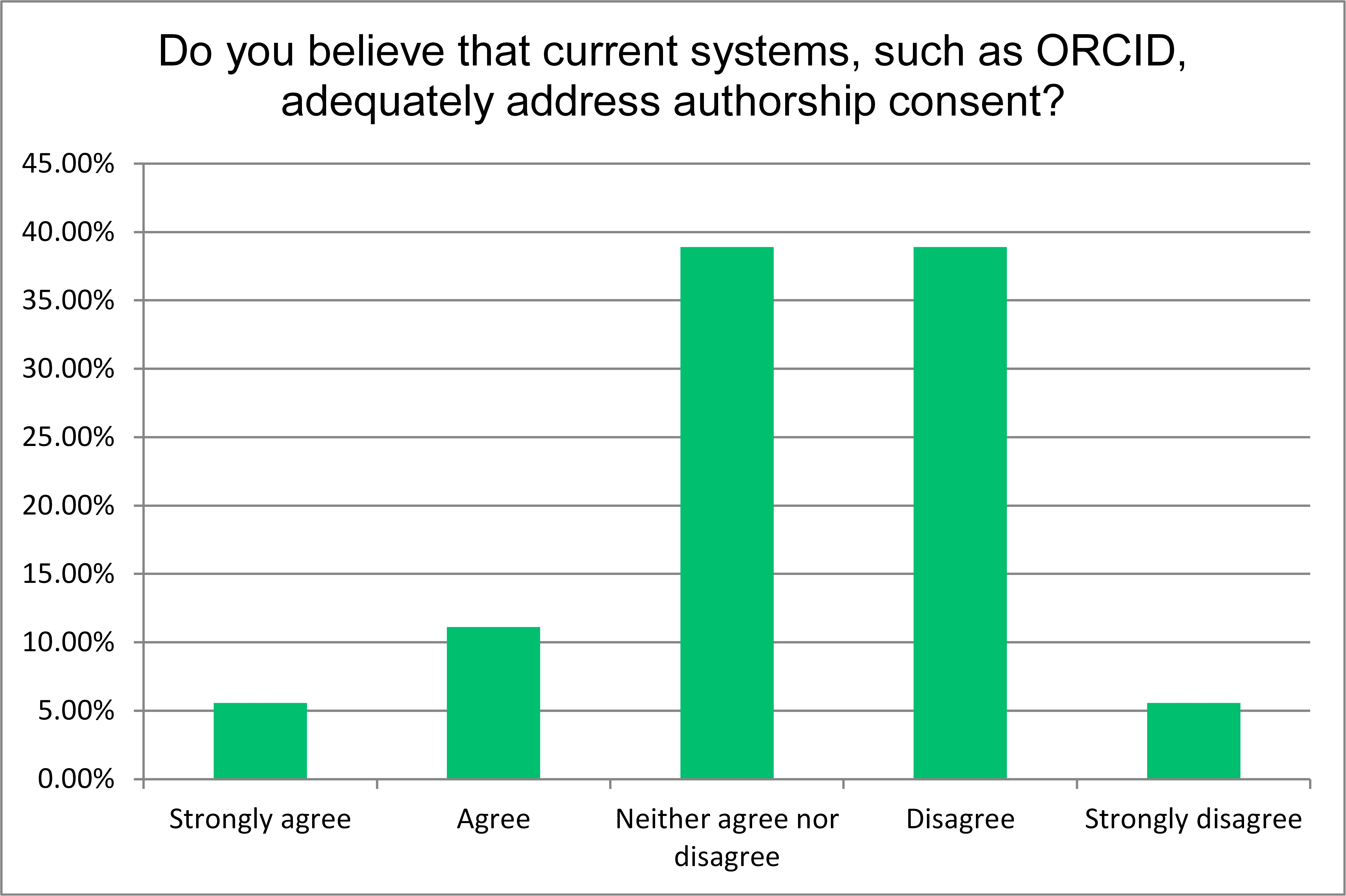}
        \caption{Perceived adequacy of ORCID in addressing authorship consent.}
        \label{fig:q12}
    \end{subfigure}
    \hfill
    \begin{subfigure}[b]{0.45\textwidth}
        \includegraphics[width=\textwidth]{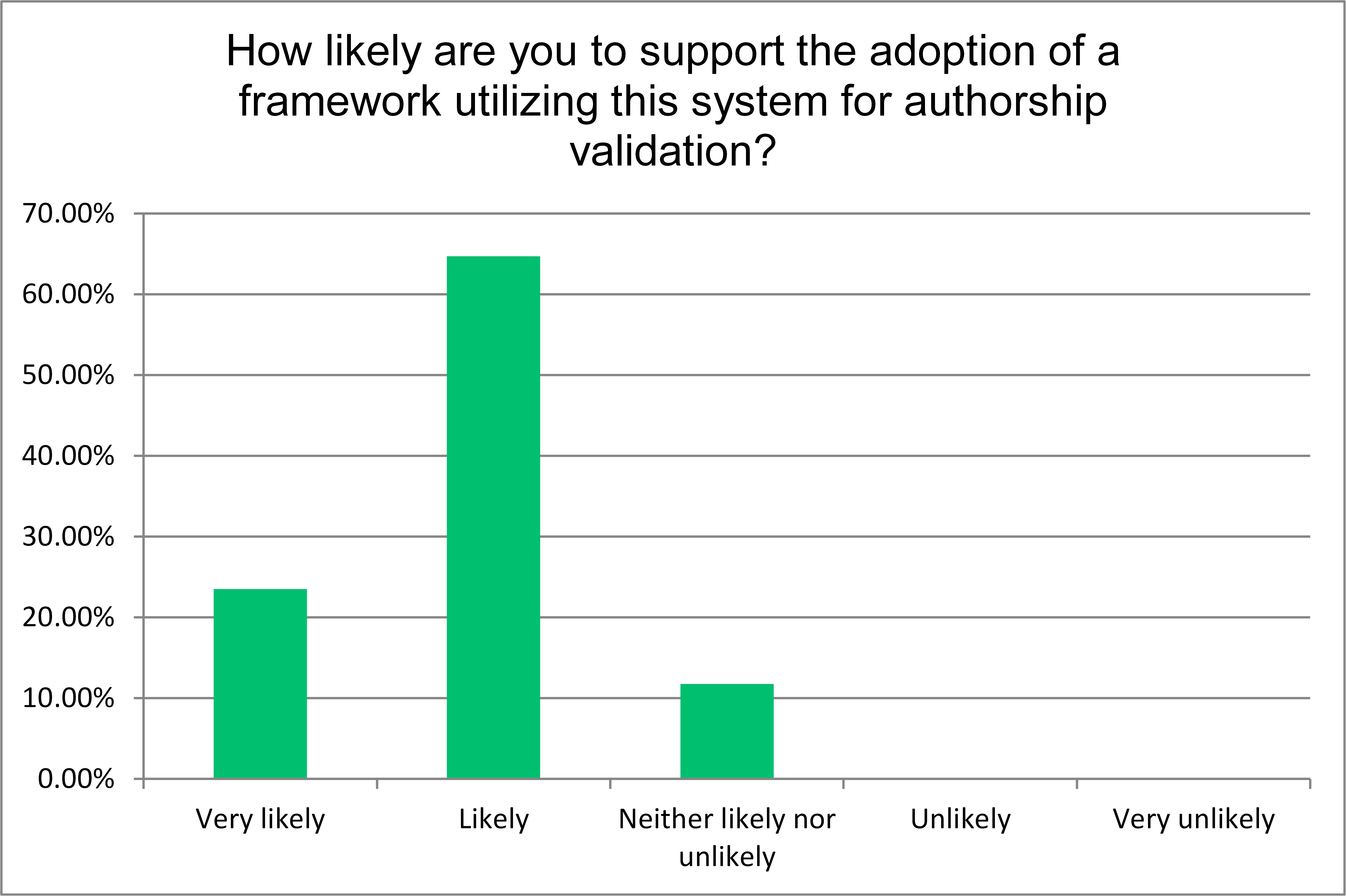}
        \caption{Support for proposed authorship validation framework.}
        \label{fig:q16}
    \end{subfigure}
    \caption{Stakeholder perceptions regarding (a) adequacy of existing solutions like ORCID, and (b) likelihood of supporting a new SSI-based authorship validation framework.}
    \label{fig:group2}
\end{figure}

\paragraph{Stakeholder Perceptions of Existing Systems}
Survey responses revealed substantial dissatisfaction with current platforms (e.g., ORCID) regarding their ability to address authorship consent effectively. Only 16.67\% of respondents agreed or strongly agreed that ORCID adequately addresses these issues, whereas 77.78\% were neutral or disagreed (Figure~\ref{fig:q12}), highlighting critical gaps that the proposed framework seeks to bridge.

\paragraph{Support for a Consent-Based Authorship Validation System}
The survey revealed strong stakeholder support for the proposed SSI- and blockchain-based framework, with 88.24\% of participants indicating likely or strong support (Figure~\ref{fig:q16}). Respondents acknowledged the need for robust, innovative solutions to improve transparency, accountability, and integrity in scholarly communication.

\begin{figure}[htbp]
    \centering
    \begin{subfigure}[b]{0.45\textwidth}
        \includegraphics[width=\textwidth]{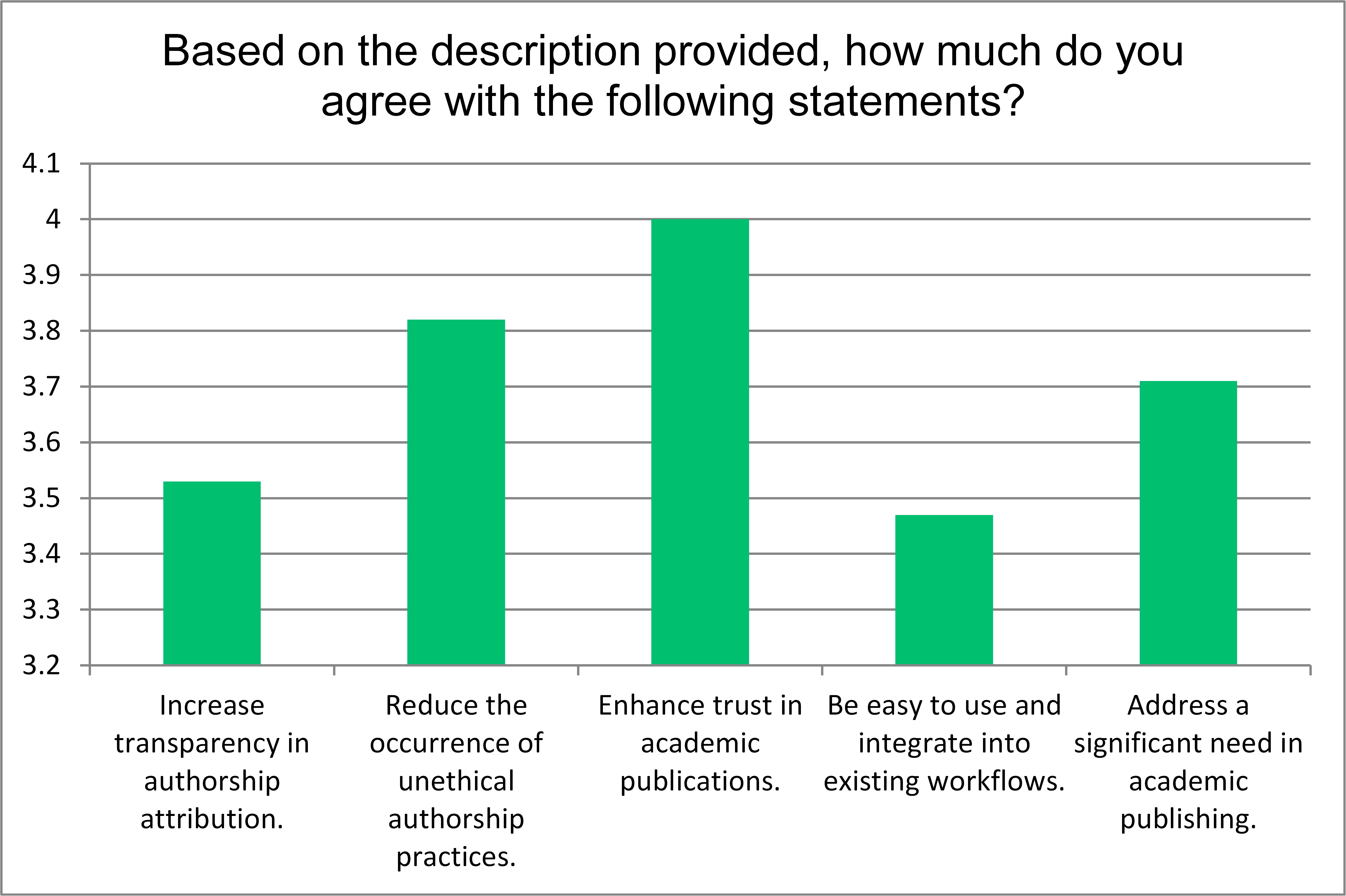}
        \caption{Perceived benefits of the proposed system.}
        \label{fig:q15}
    \end{subfigure}
    \hfill
    \begin{subfigure}[b]{0.45\textwidth}
        \includegraphics[width=\textwidth]{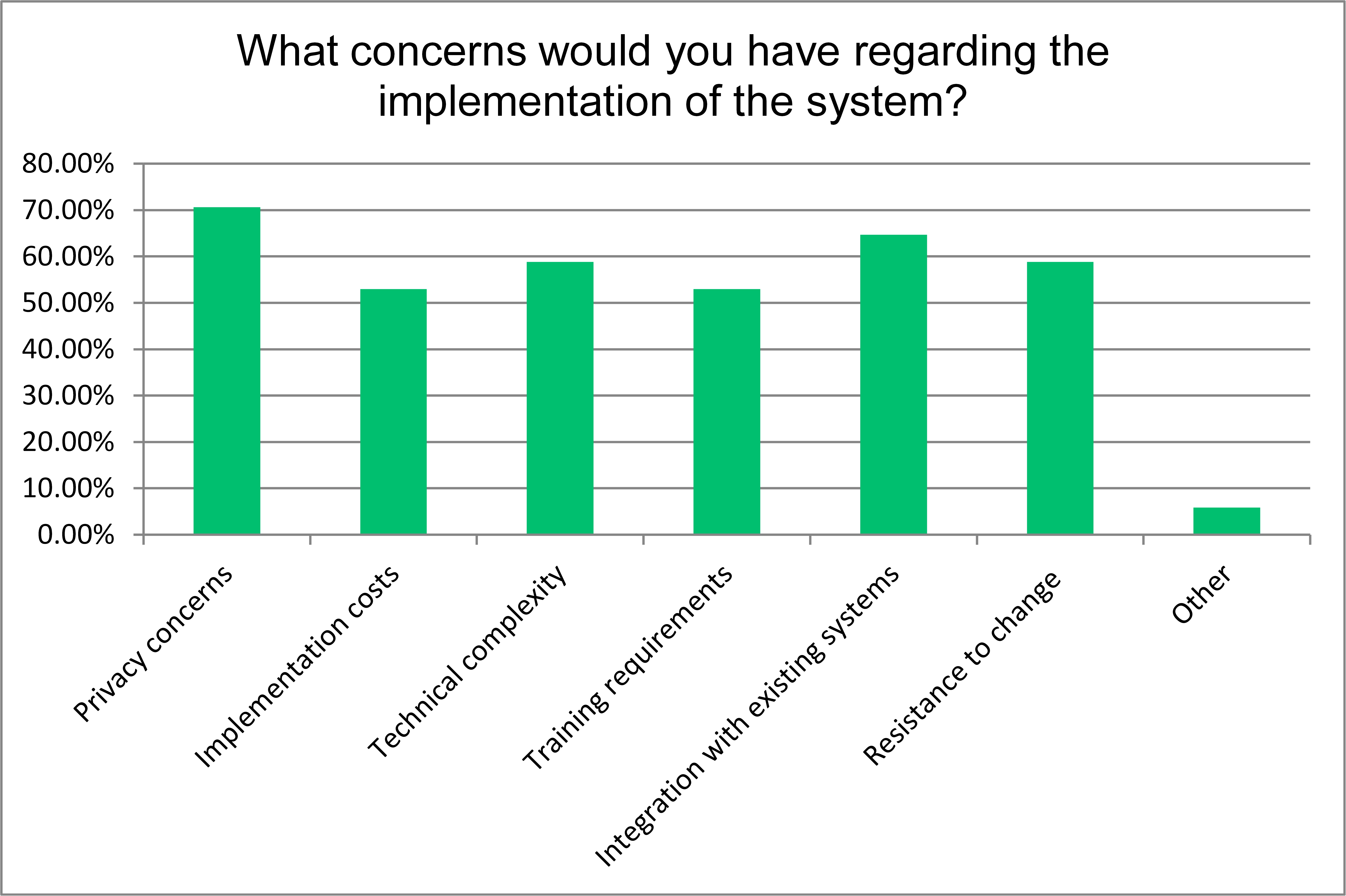}
        \caption{Concerns regarding implementation challenges.}
        \label{fig:q18}
    \end{subfigure}
    \caption{Stakeholders' (a) perceived benefits of adopting the proposed framework, and (b) primary concerns related to the implementation process.}
    \label{fig:group4}
\end{figure}

\paragraph{Concerns Regarding Implementation}
Despite widespread support, respondents identified significant implementation concerns. Notably, privacy risks (70.59\%), technical complexity (58.82\%), and integration challenges with existing academic systems (64.71\%) emerged as primary concerns (Figure~\ref{fig:q18}). These findings underscore the necessity of carefully addressing barriers to adoption.

\subsection{Implications}

The survey findings have substantial implications for the academic publishing community, particularly regarding ethical authorship validation and transparency. The prevalence of unethical practices, such as unconsented and ghost authorship, necessitates urgent solutions that ensure accountability and trust. The proposed SSI and blockchain framework directly addresses these issues by establishing tamper-proof records of authorship consent and verifiable contributions.

The high stakeholder support further indicates that the integration of automated verification and real-time notifications into publishing workflows is highly desirable. However, the concerns regarding privacy, technical complexity, and platform integration highlight critical areas that require targeted efforts. Effective implementation demands strategic investment in educational initiatives, pilot implementations, and robust infrastructure development. By proactively addressing these challenges, the proposed framework has significant potential to redefine ethical authorship practices within the scholarly ecosystem.

\subsection{Limitations}

Although the proposed framework offers robust solutions to current challenges in academic publishing, several notable limitations must be acknowledged:

\begin{itemize}
    \item \textbf{Adoption Challenges:} Stakeholder familiarity with SSI, blockchain, and advanced cryptographic techniques is limited, posing significant educational and adoption barriers.
    
    \item \textbf{Privacy Concerns:} Despite the deployment of advanced cryptographic protections, potential apprehensions regarding public Decentralized Identifiers (DIDs) and data security persist among users.
    
    \item \textbf{Interoperability and Integration:} Seamless integration of the proposed framework with existing scholarly infrastructure—such as ORCID, CrossRef, and Scopus—requires extensive technical coordination and adherence to shared standards. Without such integration, stakeholder acceptance may be limited.
\end{itemize}

These limitations underscore the necessity for carefully planned, phased implementation approaches. Strategic pilot programs, comprehensive stakeholder education, and clear interoperability guidelines will be critical to overcoming these barriers, ultimately facilitating the framework's broad adoption and effectiveness in addressing systemic authorship validation challenges.

\section{Conclusion}\label{sec:conclusion}

The integrity of academic publishing serves as a foundational pillar for trust, collaboration, and innovation across scholarly disciplines. Despite its importance, unethical authorship practices, including unconsented authorship, gift authorship, and unresolved conflicts of interest, continue to pose significant threats to scholarly credibility and trustworthiness. Although existing solutions like ORCID provide valuable mechanisms for researcher identification, they lack comprehensive capabilities for enforcing explicit consent, maintaining immutable authorship records, and ensuring robust privacy protections within peer-review processes.

To address these critical shortcomings, this paper introduced a novel decentralized framework integrating Self-Sovereign Identity (SSI) and blockchain technology. By leveraging Decentralized Identifiers (DIDs) and Verifiable Credentials (VCs), the proposed system offers secure, transparent, and decentralized mechanisms for authorship validation. Furthermore, privacy-preserving cryptographic techniques—including Zero-Knowledge Proofs (ZKPs) and selective disclosure—enable robust conflict-of-interest detection without compromising sensitive personal information. Blockchain technology was employed as a decentralized trust registry, ensuring immutability, transparency, and verifiability of authorship metadata and consent records.

The stakeholder survey results affirmed significant support for the framework's core features, particularly explicit authorship consent enforcement, automated verification, and enhanced transparency. However, respondents expressed important concerns regarding implementation barriers, notably privacy risks, technical complexity, and interoperability challenges with existing academic platforms. These insights emphasize the necessity for careful, phased implementation strategies, including pilot programs, educational initiatives, and robust infrastructure investments.

Future research directions involve expanding the framework's scope to include roles beyond authorship—such as data curators and technical contributors—to enhance inclusivity. Additionally, integrating advanced AI-driven anomaly detection methods into authorship validation processes represents another promising avenue. Such enhancements will further strengthen the model's capability to address persistent ethical challenges, fostering a more transparent, accountable, and trusted academic publishing environment.

\bibliographystyle{unsrt} 
\bibliography{main}

\begin{thebibliography}{10}

\bibitem{van2023more}
Richard Van~Noorden.
\newblock More than 10,000 research papers were retracted in 2023—a new record.
\newblock {\em Nature}, 624(7992):479--481, 2023.

\bibitem{johann2022perceptions}
Daniel Johann.
\newblock Perceptions of scientific authorship revisited: Country differences and the impact of perceived publication pressure.
\newblock {\em Science and Engineering Ethics}, 28(10), 2022.

\bibitem{shubha2021publication}
Shubha Singhal and Bhupinder~Singh Kalra.
\newblock Publication ethics: Role and responsibility of authors.
\newblock {\em Indian Journal of Gastroenterology}, 40:65--71, 2021.

\bibitem{nitin2020governing}
Nitin Naik and Paul Jenkins.
\newblock Governing principles of self-sovereign identity applied to blockchain enabled privacy preserving identity management systems.
\newblock In {\em 2020 IEEE International Symposium on Systems Engineering (ISSE)}, pages 1--6, 2020.

\bibitem{lukas2021blockchain}
Lukas Stockburger, Georgios Kokosioulis, Alivelu Mukkamala, Raghava~Rao Mukkamala, and Michel Avital.
\newblock Blockchain-enabled decentralized identity management: The case of self-sovereign identity in public transportation.
\newblock {\em Blockchain: Research and Applications}, 2(2):100014, 2021.

\bibitem{rafael2020ssibac}
Rafael Belchior, Benedikt Putz, Guenther Pernul, Miguel Correia, André Vasconcelos, and Sérgio Guerreiro.
\newblock Ssibac: Self-sovereign identity based access control.
\newblock In {\em 2020 IEEE 19th International Conference on Trust, Security and Privacy in Computing and Communications (TrustCom)}, pages 1935--1943, 2020.

\bibitem{cristian2024design}
Cristian~Nicolae Butincu and Adrian Alexandrescu.
\newblock Design aspects of decentralized identifiers and self-sovereign identity systems.
\newblock {\em IEEE Access}, 12:60928--60942, 2024.

\bibitem{alex2021selfsovereign}
Alex Preukschat and Drummond Reed.
\newblock {\em Self-Sovereign Identity: Decentralized digital identity and verifiable credentials}.
\newblock Manning, 2021.

\bibitem{stefano2023survey}
Stefano Bistarelli, Francesco Micheli, and Francesco Santini.
\newblock A survey on decentralized identifier methods for self sovereign identity.
\newblock In {\em ITASEC}, 2023.

\bibitem{efat2022dtssim}
Efat Samir, Hongyi Wu, Mohamed Azab, Chunsheng Xin, and Qiao Zhang.
\newblock Dt-ssim: A decentralized trustworthy self-sovereign identity management framework.
\newblock {\em IEEE Internet of Things Journal}, 9(11):7972--7988, 2022.

\bibitem{mohd2021blockchain}
Mohd Javaid, Abid Haleem, Ravi {Pratap Singh}, Shahbaz Khan, and Rajiv Suman.
\newblock Blockchain technology applications for industry 4.0: A literature-based review.
\newblock {\em Blockchain: Research and Applications}, 2(4):100027, 2021.

\bibitem{petr2018permissioned}
Petr Novotny, Qi~Zhang, Richard Hull, Salman Baset, Jim Laredo, Roman Vaculin, Daniel~L. Ford, Donna~N. Dillenberger, and Bonnie Lawlor.
\newblock Permissioned blockchain technologies for academic publishing.
\newblock {\em Information Services and Use}, 38(3):159--171, 2018.

\bibitem{tarkhanov2019application}
Ivan Tarkhanov, Denis Fomin-Nilov, and Michael Fomin.
\newblock Application of public blockchain to control the immutability of data in online scientific periodicals.
\newblock {\em Library Hi Tech}, 37(4):829--844, 2019.

\bibitem{mohan2019blockchain}
Vijay Mohan.
\newblock On the use of blockchain-based mechanisms to tackle academic misconduct.
\newblock {\em Research Policy}, 48(9):103805, 2019.

\bibitem{mackey2019framework}
Tim~K Mackey, Neal Shah, Ken Miyachi, James Short, and Kevin Clauson.
\newblock A framework proposal for blockchain-based scientific publishing using shared governance.
\newblock {\em Frontiers in Blockchain}, 2:19, 2019.

\bibitem{tenorio2019decentralized}
Antonio Tenorio-Forn{\'e}s, Viktor Jacynycz, David Llop-Vila, Antonio S{\'a}nchez-Ruiz, and Samer Hassan.
\newblock Towards a decentralized process for scientific publication and peer review using blockchain and ipfs.
\newblock 2019.

\bibitem{laurel2012orcid}
Laurel~L Haak, Martin Fenner, Laura Paglione, Ed~Pentz, and Howard Ratner.
\newblock Orcid: a system to uniquely identify researchers.
\newblock {\em Learned publishing}, 25(4):259--264, 2012.

\bibitem{tennant2018peerreview}
Jonathan~P Tennant.
\newblock The state of the art in peer review.
\newblock {\em FEMS Microbiology Letters}, 365(19):fny204, 08 2018.

\bibitem{jonathan2020limitations}
Jonathan~P. Tennant and Tony Ross-Hellauer.
\newblock The limitations to our understanding of peer review.
\newblock {\em Research Integrity and Peer Review}, 5(1):6, April 2020.

\bibitem{jesse2016where}
Jesse Yli-Huumo, Deokyoon Ko, Sujin Choi, Sooyong Park, and Kari Smolander.
\newblock Where is current research on blockchain technology?—a systematic review.
\newblock {\em PLOS ONE}, 11(10):1--27, 10 2016.

\bibitem{yin2023blockchain}
Xing Yin.
\newblock Blockchain technology in corporate governance: Advantages and limitations.
\newblock {\em Academic Journal of Business \& Management}, 5(11):89--103, 2023.

\bibitem{gaby2018ancile}
Gaby~G. Dagher, Jordan Mohler, Matea Milojkovic, and Praneeth~Babu Marella.
\newblock Ancile: Privacy-preserving framework for access control and interoperability of electronic health records using blockchain technology.
\newblock {\em Sustainable Cities and Society}, 39:283--297, 2018.

\bibitem{lux2020distributed}
Zoltán~András Lux, Dirk Thatmann, Sebastian Zickau, and Felix Beierle.
\newblock Distributed-ledger-based authentication with decentralized identifiers and verifiable credentials.
\newblock In {\em 2020 2nd Conference on Blockchain Research \& Applications for Innovative Networks and Services (BRAINS)}, pages 71--78, 2020.

\bibitem{Pava2023bibliometric}
Roberto~Albeiro Pava~Diaz, Rafael~Vicente Paez~Mendez, and Luis~Fernando Niño~Vasquez.
\newblock A bibliometric study of scientific production on self-sovereign identity.
\newblock {\em Ingeniería}, 28(Suppl):e19656, Feb. 2023.

\bibitem{Schardong2022self}
Frederico Schardong and Ricardo Custódio.
\newblock Self-sovereign identity: A systematic review, mapping and taxonomy.
\newblock {\em Sensors}, 22(15), 2022.

\bibitem{soltani2021survey}
Reza Soltani, Uyen~Trang Nguyen, and Aijun An.
\newblock A survey of self-sovereign identity ecosystem.
\newblock {\em Security and Communication Networks}, 2021(1):8873429, 2021.

\bibitem{yirui2022decentralized}
Yirui Bai, Hong Lei, Suozai Li, Haoyu Gao, Jun Li, and Leixiao Li.
\newblock Decentralized and self-sovereign identity in the era of blockchain: A survey.
\newblock In {\em 2022 IEEE International Conference on Blockchain (Blockchain)}, pages 500--507, 2022.

\end{thebibliography}

\end{document}